\documentclass{aastex6}
\usepackage{amssymb,amsmath}

\def\kms{km~s$^{-1}$}

\newcommand{\vsini}{\ensuremath{v_{{\mathrm e}}\sin i}}
\newcommand{\teff}{\ensuremath{T_\mathrm{eff}}}

\newcommand{\logg}{\ensuremath{\mathrm{log}\ g}}

\fullcollaborationName{The Friends of AASTeX Collaboration}

\begin{document}

%% LaTeX will automatically break titles if they run longer than
%% one line. However, you may use \\ to force a line break if
%% you desire.

\title{The chemical compositions of the 2 new HgMn stars HD 30085 and HD 30963. \\ Comparison to $\chi$ Lupi A, $\nu$ Cap and HD 174567}

%% Use \author, \affil, plus the \and command to format author and affiliation 
%% information.  If done correctly the peer review system will be able to
%% automatically put the author and affiliation information from the manuscript
%% and save the corresponding author the trouble of entering it by hand.
%%
%% The \affil should be used to document primary affiliations and the
%% \altaffil should be used for secondary affiliations, titles, or email.

%% Authors with the same affiliation can be grouped in a single
%% \author and \affil call.
\author{R. Monier\altaffilmark{1}}
\affil{LESIA, UMR 8109, Observatoire de Paris et Universit\'e Pierre et Marie Curie Sorbonne Universit\'es, place J. Janssen, Meudon, France.}

\author{E.Griffin\altaffilmark{2}}
\affil{Dominion Astrophysical Observatory, 5071 West Saanich Road, Victoria, BC, V9E 2E7, Canada}
%\email{Elizabeth.Griffin@nrc-cnrc.gc.ca}

\author{M. Gebran\altaffilmark{3}}
\affil{Department of Physics and Astronomy, Notre Dame University-Louaize, PO Box 72, Zouk Mikael, Lebanon.}

\author{T. K{\i}l{\i}\c{c}o\u{g}lu\altaffilmark{4}}
\affil{Department of Astronomy and Space Sciences, Faculty of Science, Ankara University, 06100, Turkey.}

\author{T.Merle\altaffilmark{5}}
\affil{Institut d'Astronomie et d'Astrophysique, Universit\'e Libre de
Bruxelles, CP 226, Boulevard du Triomphe, 1050 Brussels, Belgium}
%\email{tmerle@ulb.ac.be}

\and
\author{F. Royer\altaffilmark{6}}
\affil{GEPI, Observatoire de Paris, place J. Janssen, Meudon, France.}

%% Notice that each of these authors has alternate affiliations, which
%% are identified by the \altaffilmark after each name.  Specify alternate
%% affiliation information with \altaffiltext, with one command per each
%% affiliation.

%\altaffiltext{1}{AAS Journals Data Scientist}
%\altaffiltext{2}{greg.schwarz@aas.org}
%\altaffiltext{3}{AAS Journals Associate Editor-in-chief}
%\altaffiltext{4}{AAS Director of Publishing}
%\altaffiltext{5}{IOP Senior Publisher for the AAS Journals}

%% Mark off the abstract in the ``abstract'' environment.
% include old text here

\begin{abstract}
We report on a detailed abundance study of the fairly bright
slow rotators HD 30085 (A0 IV), HD 30963 (B9 III) and HD 174567 (A0 V), hitherto reported as normal stars and the sharp-lined 
$\chi$ Lupi A (B9 IV HgMn). 
In the spectra of HD 30085, HD 30963, the \ion{Hg}{2} line at 3984 \AA\ line is conspicuous and numerous lines of silicon, manganese,
chromium, titanium,  iron, strontium, yttrium and zirconium appear to be strong absorbers. 
A comparison of the mean spectra of HD 30085 and HD 30963 with a grid of synthetic spectra for selected unblended lines having reliable updated atomic data reveals large overabundances of phosphorus, titanium, chromium, manganese, strontium, yttrium, and zirconium, barium, platinum and mercury and underabundances of helium, magnesium, scandium, nickel. 
The surface abundances of $\chi$ Lupi A have been rederived on the same effective temperature scale and using the same atomic data for consistency and comparison for HD 30085 and HD 30963.
For HD 174567, milder deficiencies and excesses are found. The abundances of sodium, magnesium  and calcium have been corrected for NLTE effects. The effective temperatures, surface gravities, low projected rotational velocities and the peculiar abundance patterns of HD 30085 and HD 30963 show that these stars are 2 new HgMn stars and should be reclassified as such. HD 174567 is most likely a new marginally Chemically Peculiar star. A list of the identifications of lines absorbing more than 2 \% in the spectrum of HD 30085 is also provided.
\end{abstract}

\keywords{stars: chemically peculiar --- stars: individual (HD 30085, HD 30963, HD 174567)}

\section{Introduction}

The fairly bright stars HD 30085 (HR 1510, A0 IV, V=6.3), HD 30963 (B9 III, V=7.2) and HD 174567 (HR 7098, A0V, V=6.63) have received little attention: only 30, 10 and 47 references respectively can be found in SIMBAD for these stars. 
 We have recently undertaken a spectroscopic survey of all apparently slowly rotating bright early A stars (A0-A1V) and late B stars (B8-B9V) observable from the northern hemisphere. 
This project addresses the fundamental questions of the physics of late-B and early-A stars: i) can we find new instances of rapid rotators seen pole-on (other than Vega) and study their physical properties (gradient of temperature across the disk, limb and gravity darkening), ii) is our census of Chemically Peculiar stars complete up to the magnitude limits we adopted ? If not, what are the  physical properties of the newly found CP stars? 
HD 30085 and HD 174567 pertain to the sample of the 47 apparently slowly rotating A0-A1V stars in the northern hemisphere which satisfy \vsini $\le$ 60 \kms
and $\delta$ $\ge$ $-10^{o}$ analysed by \cite{Royer}.
An abundance analysis of selected lines outside the Balmer lines allowed to sort out these stars into 3 groups: 13 chemically peculiar stars were found (among wich 4 are new CPs), 17 superficially normal stars and 17 spectroscopic binaries (\citealt{Royer}). \cite{Monier} reported on a first analysis of the abundances of Fe, Mn and Hg and showed that HD 30085, HD 18104, HD 32867 and HD 53588 are four new HgMn stars. Similarly HD 30963 is a slowly rotating B9 star which verifies similar criteria.
\\
In this paper we report on the abundance analysis of 40 chemical elements from high resolution high quality \'echelle spectra of HD 30085, HD 30963 and HD 174567 in the optical range
and one spectrum of the HgMn star $\chi$ Lupi A ($\chi$ Lupi = HD 141556 is a double-lined spectroscopic binary, we are only interested in the abundances of $\chi$ Lupi A, the HgMn star). We compare the elemental abundances we find in these four stars to those derived for $\nu$ Cap, a bona-fide normal late B-rype star which we have recently analysed 
\citep{2018ApJ...854...50M}.
Using model atmospheres and line synthesis, we  derive for the first time the abundances of forty chemical elements in these four stars and find that they depart strongly from solar. The abundances of sodium, magnesium and calcium have been corrected for NLTE effects. The overabundances of Mn, Sr, Y, Zr, Pt and Hg and the underabundances of helium, magnesium, scandium, and nickel lead us to confirm that HD 30085 is indeed a new HgMn star and establish that HD 30963 is a new HgMn star. The superficially normal HD 174567 appears to be new mild Chemically Peculiar star. We have also reanalysed the cool HgMn star $\chi$ Lupi A A on the same temperature scale and using the same atomic data and used it as a comparison star.
\\
The paper is divided into 5 sections. The first section recapitulates what is known of HD 30085, HD 30963, HD 174567 and $\chi$ Lupi A, the second section describes new spectroscopic observations of HD 30085, HD 30963, HD 174567 and $\chi$ Lupi A. 
In the third section, we present the derivation of fundamental parameters and the spectral synthesis which we adopted to derive the abundances.
In section 4, we discuss the determination of elemental abundances for each star in light of what is known of other HgMn stars. We also provide identifications of lines absorbing more than 2 \% of the continuum in HD 30085. In the conclusion, we discuss the chemical peculiarity of the three new Chemically Peculiar stars in the light of what is known of other HgMn stars.

\section{Recap of previous spectroscopic work on HD 30085, HD 30963, HD 174567, $\nu$ Cap and $\chi$ Lupi A}

HD 30085 was ascribed a spectral type A0 IV by \cite{Cowley1969} in their survey of 1700 bright northern B9 to A9 stars with a prismatic dispersion of 125  \AA\ $mm^{-1}$ around $H_{\gamma}$. At that time, \cite{Cowley1969} did not mention any peculiarity of the spectrum. In his study of helium-weak stars, \cite{molnar72} classified HD 30085 as a B9 III from 63 \AA\ $mm^{-1}$ plates centered on  $H_{\gamma}$ using slightly different MK criteria to assign the temperature and the luminosity class than \cite{Cowley1969} did. He did not comment on any peculiarity of the lines of Si, Sr and the metals in the spectrum of HD 30085 around $H_{\gamma}$. 
HD 30085 has been detected as a fairly bright UV source with TD1 \citep{1976ubss.book.....J}, its flux steadily rising towards shorter wavelength, suggesting the star is indeed a late B star rather than an early A star. \cite{Ramella89} report on structures in the core of the \ion{Mg}{2} doublet at 4481 \AA\ from 12.4  \AA\ $mm^{-1}$ dispersion plates. 
More recently, \cite{Monier} have reported on the presence of the \ion{Hg}{2} line at 3984 \AA\ and several strong \ion{Mn}{2} in the high resolution spectra of HD 30085 and provided overabundances for Mn, Fe and Hg only 
based on the spectrum synthesis of a few lines which clearly established the HgMn nature of this star.
\\
HD 30963 was ascribed a B9 III spectral type by \cite{2010ApJ...722..605H}.
HD 174567 has been used as a "normal comparison star" by \cite{1993A&A...274..335S} in their abundance study of HgMn and superficially normal stars from IUE (International Ultraviolet Explorer) spectra.
Their modelling of coadded IUE spectra of HD 174567 revealed abundances which are nearly solar confirming the superficially normal status for this star at that time.
\\
The abundances of 22 elements in the atmosphere of $\chi$ Lupi A have been derived from optical spectra by \cite{wahlgren1994} and they are collected in Table \ref{tab:compChiLupi} .
The elemental abundances of $\nu$ Cap have been recently derived in \cite{2018ApJ...854...50M} which confirms its nearly solar abundances determined by \cite{1991MNRAS.252..116A}.

\section{Observations}

The high resolution spectra of HD 30085, HD 30963 and HD 174567 have been obtained 
at Observatoire de Haute Provence with SOPHIE, the \'echelle spectrograph in its high resolution mode (R=75000) yielding a full spectral coverage from 3820 \AA\ to 6930 \AA\ in 39 orders.
A detailed technical description of SOPHIE is given in \cite{Perruchot}. SOPHIE is a cross-dispersed, environmentally stabilized \'echelle spectrograph dedicated to high precision radial velocity. The spectra are extracted online from the detector images using a specific pipeline adapted from that of HARPS
(High Accuracy Radial Velocity Planet Searcher \footnote{https://www.eso.org/sci/facilities/lasilla/instruments/harps.html})
. The sequence or reductions starts with the localization of the 39 orders on the 2D images, the optimal extraction of each order, the wavelength calibration and finally the correction for  flat-field producing a two dimensional spectrum (e2ds). 
 We have normalised each reduced order separately using a Chebychev polynomial fit with sigma clipping, rejecting points above or below 1 $\sigma$ of the local continuum. Normalized orders were then  merged together, corrected  by the blaze function and resampled into a constant wavelength step of about 0.02\,\AA\ \cite[see][for more details]{Royer}.
The observing dates, exposure times and Signal to Noise ratios achieved for each
observing run are collected in Table \ref{tab:obs}.
 For HD 30085 and HD 30963, the comparison of individual spectra taken at different epochs did not reveal any convincing radial velocity nor residual flux variations in the lines. We therefore coadded the individual orders into merged orders to enhance the signal-to-noise ratio up to 350.
Complementary monoorder spectra of HD 30085 have been obtained at Dominion Astrophysical Observatory by EG with a  lower resolving power R=45000. The three wavelengths intervals observed span
from 3874 \AA\ to 4020 \AA,  from 4038 \AA\ to 4183 \AA\ and from 4430 \AA\ to 4560 \AA. The  $H_{\epsilon}$  and $H_{\delta}$ regions were observed in order to validate the fundamental parameters derived from the Str\"omgren's photometry (see section 4.1) and confirm the presence of the \ion{Hg}{2} line at 3983.93 \AA.

\begin{table}
\caption{Log of the observed and archival spectra}
\label{tab:obs}
\centering
\begin{tabular}{||c|c|c|c|c|c|c|}
\hline  \hline
Star ID   &  Date & instrument & R & Barycentric Julian Date & duration (s)& S/N   \\ \hline
HD 30085  & 13/02/2012   & SOPHIE       & 75000	 & 2455971.36664683 &   800 &  216  	      \\ 
HD 30085  & 10/12/2013   & SOPHIE       & 75000	 & 2456637.61970650 &  1200 &  269  	      \\
HD 30085 & 07/02/2013 & DAO & 45000 & 2456331.255 & 3600 &  220\\
HD 30085 & 02/03/2013 & DAO & 45000 & 2456354.176 & 3600 & 220  \\
HD 30085 & 03/03/2013 & DAO & 45000 & 2456355.136 & 3600 & 220 \\
HD 30085 & 18/08/2015 & DAO & 45000 & 2457253.503 & 2500 & 250 \\
HD 30085 & 19/08/2015 & DAO & 45000 & 2457254.507 & 2400 & 250 \\
HD 30085 & 10/09/2015 & DAO & 45000 & 2457276.500 & 2100 & 240 \\
HD 30085 & 11/09/2015 & DAO & 45000 & 2457277.478 & 3600 & 220 \\
HD 30085 & 12/09/2015 & DAO & 45000 & 2457278.435 & 2700 & 250\\
HD 30085 & 21/10/2015 & DAO & 45000 & 2457317.459 & 2700 & 250 \\
HD 30085 & 24/11/2015 & DAO & 45000 & 2457351.233& 3600 & 220 \\
HD 30085 & 24/11/2015 & DAO & 45000 & 2457351.261& 1200 & 240 \\
HD 30085 & 24/11/2015 & DAO & 45000 & 2457351.506 & 1800 & 240 \\
HD 30085 & 24/11/2015 & DAO & 45000 & 2457351.527 & 1800 & 240 \\
HD 30085 & 25/11/2015 & DAO & 45000 & 2457352.274& 3600 & 220 \\
HD 30085 & 25/11/2015 & DAO & 45000 & 2457352.501 & 3600 & 220\\
HD 30085 & 26/11/2015 & DAO & 45000 & 2457353.222 & 3600 & 220 \\
HD 30085 & 26/11/2015 & DAO & 45000 & 2457353.476 & 3600 & 220 \\
HD 30085 & 27/11/2015 & DAO & 45000 & 2457354.225 & 3600 & 220 \\
HD 30085 & 25/12/2015 & DAO & 45000 & 2457382.278 & 3600 & 220 \\
HD 30085 & 30/12/2015 & DAO & 45000 & 2457387.317 & 3600 & 220 \\
HD 30085 & 31/12/2015 & DAO & 45000 & 2457388.274 & 3600 & 220 \\
HD 30085 & 01/01/2016 & DAO & 45000 & 2457389.247 & 3600 & 220 \\
HD 30085 & 23/01/2016 & DAO & 45000 & 2457411.159 & 3600 & 220 \\
HD 30963  & 28/11/2015   & SOPHIE    & 75000     & 2457354.49106071      & 2400   & 118              \\
HD 30963  & 28/11/2015   & SOPHIE    & 75000     & 2457355.48326524       & 1800   & 78              \\
HD 30963  & 30/11/2015   & SOPHIE    & 75000     & 2457356.51212353      & 1800  & 140             \\
HD 30963  & 30/11/2015   & SOPHIE    & 75000     & 2457357.46945236      & 1800   & 101              \\
HD 30963  & 01/12/2015   & SOPHIE    & 75000     & 2457358.43199975     & 1680  & 167              \\
HD 30963  & 13/12/2016   & SOPHIE    & 75000     & 2457736.46584169    & 900   & 129              \\
HD 30963  & 13/12/2016   & SOPHIE    & 75000     & 2457736.49875775                  & 900   & 123              \\
HD 30963  & 14/12/2016   & SOPHIE    & 75000     & 2457737.44071886                  & 1800   & 165             \\ 
HD 174567 & 05/08/2009   & SOPHIE    & 75000     & 2455049.36018803     & 1200   & 224             \\
HD 141556 & 04/02/2013   & FEROS   & 48000      & 2456327.37979         & 50    &   325         \\ 
%HD 141556 &              & UVES      & 107200 &          &     &            \\
%HD 141556 &              & HARPSPOL  I &      &          &     &          \\   
\hline		
\hline	
\end{tabular}
\end{table}

\section{Model atmosphere analysis and synthetic spectra computation}

The abundances of forty chemical elements have been derived by iteratively
adjusting synthetic spectra to the normalized spectra and looking for the best fit to carefully selected unblended lines.
Specifically, synthetic spectra were computed assuming
LTE using \cite{Hubeny} SYNSPEC49 code which calculates lines for
elements up to Z=99. For selected lines of \ion{Na}{1}, \ion{Mg}{1} and \ion{Ca}{2}, we have provided Non-LTE 
abundances.

\subsection{Fundamental parameters}

The fundamental data for HD 30085, HD 30963, $\nu$ Cap, HD 174567, and HD 141556 are collected in Table \ref{tab:HD30085data}.  
The spectral type retrieved from SIMBAD appears in column 2 and the apparent magnitudes
in column 3, the Str\"omgren indexes $b -y$, $m_{1}$ and $c_{1}$ in columns 4, 5, 6. 
The photometric data was taken from \cite{1998A&AS..129..431H}.
% moved from preceeding paragraph
The radial velocity of HD 30085, HD 174567 and $\nu$ Cap are those derived in \cite{Royer} using cross-correlation techniques, avoiding the Balmer lines and the atmospheric telluric lines. The normalized spectrum was cross-correlated with a synthetic template extracted from the POLLUX database\footnote{\url{http://pollux.graal.univ-montp2.fr}}  \citep{Palacios} corresponding to the parameters $T_\mathrm{eff}=11000$\,K, $\log g=4$ and solar abundances. A parabolic fit of the upper part of the resulting cross-correlation function yields the Doppler shift, which is then used to shift spectra to rest wavelengths. The projected rotational velocities were derived from the position of the first zero of the Fourier transform of individual lines, they are taken from \cite{Royer}. The radial velocity and projected equatorial velocity of HD 30085, HD 30963, $\nu$ Cap, HD 174567 and HD 141556 are collected in Table 2. 
\\
For the five stars, the effective temperature (\teff)  and surface gravity (\logg) were determined  using the UVBYBETA code developed by \cite{Napiwotzki} and appear in columns 7 and 8. This code is based on the \cite{1985MNRAS.217..305M} grid, which calibrates the $uvby\beta$ photometry in terms of \teff \ and \logg.  
The estimated errors on \teff \ and \logg, are $\pm$125 K and $\pm$0.20 dex, respectively (see Sec. 4.2 in Napiwotzki et al. 1993). 
% According to Crawford (****), the total extinction $A_{\nu} = 0$ in the direction of HD 30085, we therefore did not correct the colors for any extinction.\\

% Table fundamental parameters 

\begin{table}
\caption{Adopted fundamental parameters for HD\,30085, HD\,30963, 
$\nu$\,Cap, HD\,174567 and HD\,141556}
\label{tab:HD30085data}
\centering
\begin{tabular}{||ll|c|c|c|c|c|c|c|c|c|c|c||}
\hline  \hline
  & & HD\,30085 & HD\,30963 & $\nu$\,Cap & HD\,174567 & HD\,141556 \\
  \hline
  Sp.T. & & A0\,IV & B9\,III & B9\,IV & A0\,V & B9\,IV \\
  V & & 6.37 & 7.20 & 4.76 & 6.63 & 3.95 \\
  $b-y$ & & -0.038 & -0.021 & -0.021 & 0.008 & -0.020 \\
  $m_1$ & & 0.135 & 0.134 & 0.134 & 0.123 & 0.129 \\
  $c_1$ & & 0.849 & 1.015 & 1.015 & 1.083 & 0.948 \\
  $T_{\rm{eff}}$ & [K] & $11300\pm200$ & $11476\pm200$ & $10300\pm200$ & 
$10200\pm200$ & $10608\pm200$ \\
  log\,$g$ & (cgs) & $3.95\pm0.20$ & $3.66\pm0.20$ & $3.90\pm0.20$ & 
$3.55\pm0.20$ & $3.98\pm0.20$ \\
  $v_{\rm{e}}$\,sin$i$ & [km\,s$^{-1}$] & 26.0 & 37.0 & 24.0 & 10.5 & 2.0 \\
  $v_{\rm{micr.}}$ & [km\,s$^{-1}$] & $0.0^{+0.4}_{-0.0}$ & 
$0.1^{+0.3}_{-0.1}$ & 0.50 & $0.95\pm0.15$ & 0.10 \\
  $v_{\rm{rad}}$ & [km\,s$^{-1}$] & 8.27 & 3.52 & -11.39 & -10.84 & 15.30 \\
  $M$ & [$M_\odot$] & $3.1\pm0.4$ & $3.7\pm0.4$ & $2.8\pm0.3$ & 
$3.4\pm0.4$ & $2.8\pm0.3$ \\
  log\,$t$ & ($t$ in yr.) & $8.35\pm0.05$ & $8.27\pm0.07$ & 
$8.50\pm0.06$ & $8.37\pm0.15$ & $8.46\pm0.10$\\
  $t_{MS}$ & (\%) & 64 & 85 & 69 & 86 & 63 \\
  \hline \hline
\end{tabular}
\end{table}

\subsubsection{Microturbulent velocity determinations}

\begin{figure}[h!]
\vskip 0.5cm
   \centering
      \includegraphics[scale=0.90]{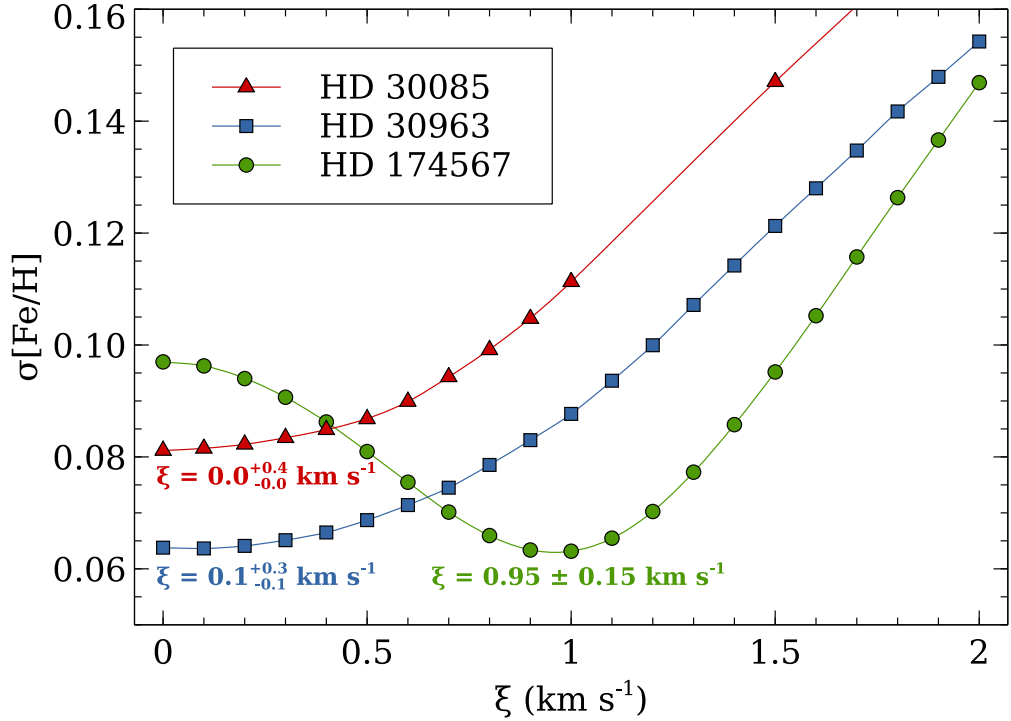}
   \caption{Microturbulent velocity determinations for HD 30085, HD 30963 and HD 174567}
   \label{figure1}\vskip 0.5cm
\end{figure}

In order to derive the microturbulent velocity of HD 30085, HD 30963 and HD 174567, we have simultaneously derived the iron abundance
[Fe/H] for fifty unblended \ion{Fe}{2} lines and  a set of microturbulent
velocities ranging from 0.0 to 2.0 \kms. Figure \,\ref{figure1} shows the standard
deviation of the derived [Fe/H] as a function of the microturbulent velocity.
The adopted microturbulent velocities are the values which minimize the standard deviations ie. for that value, all \ion{Fe}{2} lines yield the same iron abundance.
The microturbulent velocities of the three stars are collected in Table \ref{tab:HD30085data}.

% microturbulent velocity plot to include here 

\subsubsection{Location of the five stars in a \logg,$\log$ \teff diagram}

\begin{figure}[h!]
%\vskip 0.5cm
    \centering
      \includegraphics[scale=0.90]{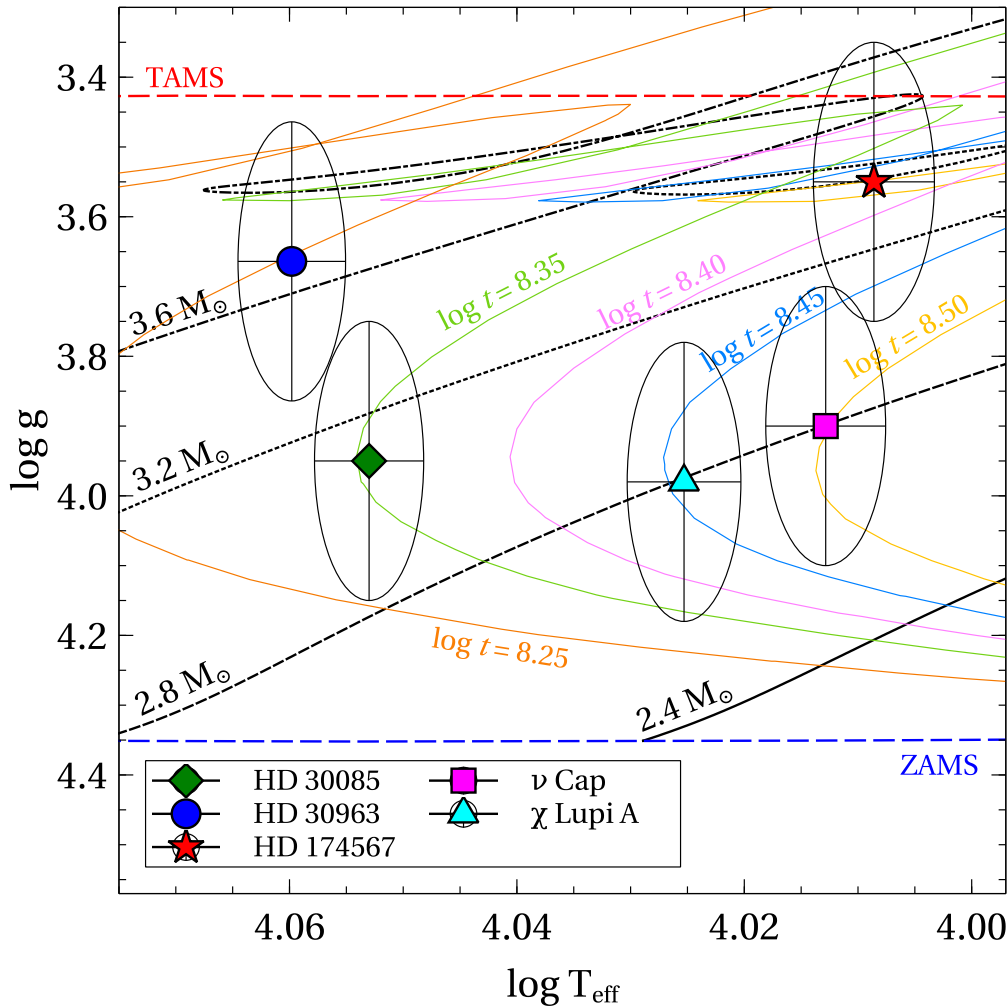}
      \caption{Location of HD 30085, HD 30963, $\chi$ Lupi A, $\nu$ Cap and HD 174567 in a gravity-temperature diagram.
     \label{figure2}}\vskip 0.5cm
\end{figure}

The locations of HD\,30085, HD\,30963, HD\,174567, $\nu$\,Cap, and 
$\chi$\,Lupi in a theoretical gravity-temperature (log\,$g$, 
log\,$T_{\mathrm{eff}}$) diagram are shown in Figure \ref{figure2}. 
Evolutionary tracks and isochrones from 
\citet{bressanetal12}\footnote{http://stev.oapd.inaf.it/cgi-bin/cmd} are 
displayed for masses from 2.4 to 3.6 $M_{\odot}$ with a step of 0.4 
$M_{\odot}$ and for log\,$t$ (where $t$ is in years) of 8.25, 8.35, 8.40, 
8.45, and 8.50. The evolutionary tracks are computed for a solar initial 
metallicity Z=0.017 including microscopic diffusion. The isocrones are 
retrieved for the current solar composition Z=0.0152. From this diagram, 
we have estimated masses, ages and fractional lifetimes on the Main 
Sequence which are collected in the last three rows of Table 
\ref{tab:HD30085data}.

% Table fundamental parameters 

% microturbulent velocity plot to include here 

\subsection{Model atmospheres}

The ATLAS9 code \citep{Kurucz} was used to compute a first model atmosphere for the effective temperature and surface gravity of each star assuming
a plane parallel geometry, a gas in hydrostatic and radiative equilibrium and local thermodynamical equilibrium. The ATLAS9 model atmosphere contains 72 layers with a regular increase in
$\log \tau_{Ross} = 0.125$ and was calculated assuming a solar chemical composition \citep{1998SSRv...85..161G}. It was converged up to $\log \tau = -5.00$ in order to attempt reproduce the cores of the Balmer lines.
This ATLAS9 version uses the new opacity
distribution function (ODF) of \cite{Castelli2003} computed for that solar
chemical composition.
Once a first set of elemental abundances was derived using the ATLAS9 model atmosphere, the atmospheric structure was recomputed for these abundances using the Opacity sampling ATLAS12 code \citep{atlas12,atlas12_2}.
New slightly different abundances were then derived and a new ATLAS12 model recomputed until the abundances of iteration (n-1) differed of those of iteration (n) by less than $\pm$ 0.10 dex.

\subsection{The linelist}

Atomic linelists from Kurucz's database have provided the basis to construct our linelist\footnote{http://kurucz.harvard.edu/lineslists/ }. These lists collect data mostly from the literature for light and heavy elements (usually critically evaluated transition probabilities from \cite{Martin88} and \cite{Fuhr1988}) and computed data by \cite{Kurucz} for the iron group elements.
A first linelist was built from Kurucz's gfall21oct16.dat\footnote{\url{http://kurucz.harvard.edu/linelists/gfnew/gfall21oct16.dat}} which includes hyperfine splitting levels.
We then updated several oscillator strengths and damping parameters with more recent and accurate determinations when necessary.
As a rule, we have preferred NIST\footnote{http://physics.nist.gov/cgi-bin/AtData/linesform} and \cite{W96} oscillator strengths for CNO and also \cite{2006A&A...445.1165N}. 
The H I lines are calculated using \cite{1973ApJS...25...37V} tables upgraded by 
Schoening \& Butler up to H 10.
The He I lines are computed in SYNSPEC49 by using 
specific tables, either from \cite{shamey} or \cite{Dimitrijevic1984}.
%
%For a few elements deemed to be important, we resorted to specific publications: Si II \citep{1981A&AS...44..171A,artru1986}, Ti II \citep{Pickering2002}, Cr I \citep{Sobeck2007}, Cr II %\citep{1990MNRAS.247..611S,2012ApJS..202...15S,2014ApJS..213...28S}, Mn II \citep{2013ApJS..205...14K},
%Ga II \citep{1994AZh....71...83R,2000A&A...363..815N}, Y II \citep{1990asos.conf..200N}, Ba II (Curry, 2004). 

We introduced hyperfine components for a few lines of Mn II \citep{1999MNRAS.306..107H}, isotopic and hyperfine transitions for a few lines of \ion{Ga}{2} \citep{2000A&A...363..815N}.
% Hg li linelist here
To model the \ion{Hg}{2} line at 3983.93 \AA, we have included 9 hyperfine transitions from the various isotopes from $\mathrm{Hg}^{196}$ up to $\mathrm{Hg}^{204}$ from \cite{Do03}. 
% For \ion{Pt}{1}, the linelist provided in Den Hartog et al. (2005) was used to upgrade older publications.

For the Rare Earths, we retrieved all relevant transitions from the DREAM database\footnote{http://www.umh.ac.be/astro/dream.shtml}.
We also used specific publications reporting on laboratory measurements: Sm II \citep{LA06}, Nd II \citep{2003ApJS..148..543D}, Eu II \citep{LWHS}, Tb II \citep{law01}.
In Kurucz's linelists, the damping constants are taken from the literature when available. For the iron group elements, they come from Kurucz's (1992) computations. Additional damping constansts for a few Si II lines were taken from \cite{1988A&A...192..249L}. When they are not available from the linelist, damping constants are actually computed in SYNSPEC49 using an approximate formula \citep{1981SAOSR.391.....K}.

\subsection{Spectrum synthesis}

We have used only unblended lines to derive the abundances.
For a given element, the final abundance is a weighted mean of the abundances derived for each transition (the weights are derived from the NIST grade assigned to that particular transition).
For several elements, in particular the heaviest elements, only one unblended line was available and the abundance should be regarded as uncertain.
For each modeled transition, the adopted abundance is that which provides the best fit 
calculated with SYNSPEC49 to the observed normalized profile.
A grid of synthetic spectra was computed with SYNSPEC49 \citep{Hubeny} to model each selected unblended
 line of the forty elements  for the four stars. Computations were iterated varying the unknown abundance until minimization of the chi-square between the observed and synthetic spectrum was achieved.
For HD 30085, the final synthetic spectrum has allowed the identification of most observed lines. The identifications of lines absorbing more than 2\% of the continuum are collected in Table \ref{tab:ident}.

\section{Abundance determinations}

We discuss here the identifications of chemical elements, the lines selected for abundance determinations  and the abundance determinations for each element. The abundances derived for each transition and the final abundances
\footnote{We here refer to the absolute abundance in the star:
$\log_{10}\left({\mathrm{X}\over\mathrm{H}}\right)_{\star}$ where $\log_{10}\left(\mathrm{H} \right)=12.0$}
are collected with their errors in Table \ref{tab:abundAll}.
 We also provide non-LTE abundance corrections for three light
elements, namely \ion{Na}{1}, \ion{Mg}{1} and \ion{Ca}{2}. 
We use the Formato code (Merle et al., in prep.) to build simple model
atoms, using atomic data from NIST \citep{Kramida2018} for energy
levels, from VALD \citep{rya2015} for radiative bound-bound
transitions and from TOPbase \citep{1992RMxAA..23..107C} for radiative
bound-free transitions. For inelastic collisions with electrons, we use
the semi-empirical formula from \cite{seaton62} for which a radiative
counterpart exist and a collision strength of one for transitions with a
forbidden radiative couterpart. For inelastic collisions with hydrogen,
we use the formula from \cite{drawin1969} without scaling factor, excepted
for \ion{Mg}{1} where we implemented the mechanical quantum data from
\cite{GUITOU201594}.
The statistical equilibrium and the radiative
transfer equation for each level and line in each model atom are
consistently solved using the non-LTE 1D radiative transfer code MULTI
\citep{Carlsson86,Carlsson92}.

\subsection{Helium}

\begin{figure}[h!]
\vskip 0.5cm
   \centering
      \includegraphics[scale=0.90]{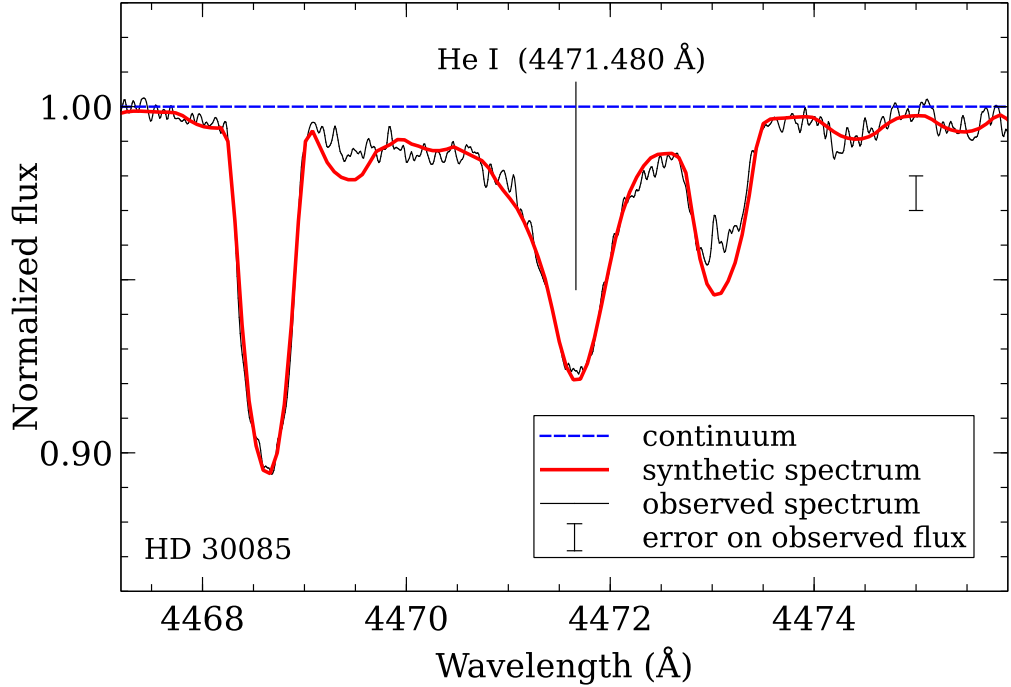}
   \caption{Synthesis of the \ion{He}{1} line at 4471.480 \AA\ for HD 30085.}
   \label{figure3}
\end{figure}

The strongest helium lines unambiguously detected are: $\lambda$  4026.19 \AA, $\lambda$ 4471.48 \AA, $\lambda$ 5875.621 \AA. The others are either weak or embedded into blends.
The least blended, $\lambda$ 4471.48 \AA, has been synthesized to derive the helium LTE abundance. Helium is underabundant in all stars, the defficiency is more important in the three HgMn stars.
Figure \,\ref{figure3} displays the best fit achieved to model the \ion{He}{1} line at 4471.480 \AA\ for HD 30085. 

% refer to He I line synthesis figure here 

\subsection{The light elements (C to Ti)}

\subsubsection{Carbon}

Most of the expected \ion{C}{2} lines are blended with more abundant species. The least blended is the high excitation \ion{C}{2} triplet at 4267.26 \AA\ which coincides with a  5 \% line at 4267.18 \AA\ in HD 30085 for instance.
Carbon is underabundant in HD 30085, HD 174567 and $\chi$ Lupi A but solar in HD 30963. \cite{1987SvA....31..151S} has shown that the \ion{C}{2} line at 4267.26 \AA\ is prone to NLTE effects for B stars hotter than 15000 K. He suggests to use the \ion{C}{2} lines at 3919 and 3921 \AA\ less influenced by departures from LTE to derive carbon abundances. These lines yield the same carbon abundances.

\subsubsection{Oxygen}

For \ion{O}{1}, the strongest expected allowed lines are those of multiplet 7 (NIST quality B+). The nine transitions of multiplet 7 dominate the opacity from 6155 \AA\ to 6159 \AA\. 
Other lines of \ion{O}{1} are blended with lines of iron-peak elements and were therefore discarded to derive the oxygen abundance.
Oxygen is underabundant in HD 30085 and $\chi$ Lupi A, solar in HD 30963 and slightly overabundant in HD 174567.

\subsubsection{Sodium}

We have used the 2 \ion{Na}{1} lines at 4497.66 \AA\ (NIST quality B+) and the resonance \ion{Na}{1} lines at 5889.95 and 5895.92 \AA\ (quality AA). The LTE overabundances of sodium derived from these lines are -4.97 for HD 30085 and -5.67 (ie. solar) upper limit for HD 30963 and HD 174567. The NLTE corrections for the resonance \ion{Na}{1} lines at 5889.95 and 5895.92 \AA\ are about -0.52 and -0.60 dex respectively which yields Non-LTE abundances of Na of -5.61 in HD 30085 and -6.27 for HD 30963 and HD 174567 and -5.57 for $\chi$ Lupi A, well below the LTE determinations. Sodium is underabundant in HD 30963 and HD 174567 and slightly overabundant in HD 30085 and $\chi$ Lupi A.

\subsubsection{Magnesium}

The unblended \ion{Mg}{2} lines at  4390.514 \AA\ and 4390.572 \AA\ yield LTE abundances of -4.64, -4.57, -4.42 and -4.46  for HD 30085, HD 30963 and HD 174567 and $\chi$ Lupi A respectively.
The unblended \ion{Mg}{1} line at 5172.68 \AA\ yields similar LTE abundances.
We adopt the abundance derived from the unblended \ion{Mg}{1} line at 5172.68 \AA\ corrected with the NLTE correction of about -0.33 dex, which yields -4.42, -4.85, -4.59
and -5.02 respectively. Magnesium therefore appears to be underabundant in each star.

\subsubsection{Aluminium}

The synthesis of the unblended  \ion{Al}{2} line at 4663.056 \AA\ shows that aluminium is severely depleted in the three Chemically Peculiar stars but not in the superficially normal HD 174567.

\subsubsection{Silicon}

The synthesis of strong and unblended lines of \ion{Si}{2} shows that silicon is slightly overabundant in HD 30085, HD 30963 and HD 174567 and slightly depleted in $\chi$ Lupi A.

\subsubsection{Phosphorus}

The lines of \ion{P}{2} expected strongest are $\lambda$ 6024.18 \AA\ and 6043.12 \AA\ of multiplet 2. Phosphorus is overabundant in the four stars.

\subsubsection{Sulfur}

 The lines of \ion{S}{2} at 4162.67 \AA, 4142.25 \AA\ and 4153.06 \AA\ of multiplet 4 are clearly present and unblended in the spectra of all stars. 
Sulfur is solar in HD 30085, overabundant in HD 30963 and $\chi$ Lupi A and underabundant in HD 30963.

\subsubsection{Calcium}

The synthesis of the unblended \ion{Ca}{2} lines at 3933.66 \AA\ and 5019.97 \AA\ yields the calcium LTE abundances, respectively -5.64 , -5.46, -5.64 and -5.83 in HD 30085, HD 30963, HD 174567 and $\chi$ Lupi A. The Non-LTE correction for the 3933.66 \AA\ amounts to +0.37 dex, yielding final overabundances of -5.22 for HD 30085 and HD 174567 and -5.37 for HD 30963 and -5.46 for $\chi$ Lupi A respectively. Calcium therefore appears to be overabundant in all four stars.

\subsubsection{Scandium}

The line of multiplet 1 of \ion{Sc}{2} at 4246.820 \AA\ is the only unblended line (quality A' in NIST). Scandium appears to be overabundant in HD 30085, HD 30963 and HD 174567 but underabundant in $\chi$ Lupi A

\subsubsection{Titanium}

We have used ten unblended lines of \ion{Ti}{2}  listed in Table \ref{tab:abundAll} to derive the mean titanium abundances. Titanium is overabundant in all four stars.

\subsection{The iron-peak elements (V to Zn) and gallium}

\subsubsection{Vanadium}

The vanadium abundance has been estimated using three lines of \ion{V}{2} at  4005.705 \AA, 4023.378 \AA\ and 4036.78 \AA. The adjustment of these lines suggests a significant underabundance of vanadium in HD 30085, overabundances for HD 30963 and HD 174567 and a nearly solar abundance for $\chi$ Lupi A.

\subsubsection{Chromium}

Several strong and unblended lines of \ion{Cr}{2} could easily be found in the spectra of the four stars. Six unblended lines of \ion{Cr}{2}  were synthesized and yield chromium overabundances in the four stars. The three resonance lines of \ion{Cr}{1} lines at 4245.35, 4274.82 and 4289.73 \AA\ are also present and yield lower overabundances. The next lines of \ion{Cr}{1} expected to contribute opacity are those of multiplet 3 but they are all blended. Lines of higher multiplets are expected too faint to be detectable even at an overabundance of -5.43. We have retained the overabundances from \ion{Cr}{2} which is the dominant ionisation stage. 

\subsubsection{Manganese}

The lines of \ion{Mn}{2} are conspicuous in the spectra of HD 30085 and HD 30963. Two lines have hyperfine structures published by \cite{1999MNRAS.306..107H}. They yield overabundances in all stars, especially large for HD 30085 and HD 30963.
\\
The resonance lines of \ion{Mn}{1} at 4030.753 \AA\ and 4034.483 \AA\ are properly reproduced by a lower manganese overabundance of about -5.13 for HD 30085 for instance. Three other lines at 4055.544 \AA\ and 4526.530 \AA\ and 4823.524 \AA\ absorb a few \% of the continuum and are properly reproduced by the same abundance. The other Mn I lines are very weak and blended.

\subsubsection{Iron}

%Fifty lines of \ion{Fe}{2} of multiplet 27 up to 74 are easily found, most of them are moderately strong features (5 to 20 \% absorption). 
The abundance of iron was derived from the eleven \ion{Fe}{2} lines, all of which are sensitive to small changes in $[\frac{Fe}{H}]$. Iron is overabundant in HD 30085, HD 30963 and HD 174567 and slightly underabundant for $\chi$ Lupi A.
\\
The ten lines of \ion{Fe}{1} listed in \cite{2004A&A...425..263C} are present, the strongest are 4045.812 \AA\ and 4383.545 \AA\ and are blended in all stars. The others range from 2 to 5 \% absorption and are often blended. The only unblended \ion{Fe}{1} line is 4383.545 \AA\ , it is adjusted with a solar abundance of iron for all stars.

\subsubsection{Nickel}

The \ion{Ni}{2} line at $\lambda$ 4067.031 \AA\ is weak in the spectra of HD 30085 and $\chi$ Lupi A, absorbing about 1\% of the continuum and is properly fit with a pronounced underabundance of nickel in HD 30085 and in $\chi$ Lupi A. In HD 30963 and HD 174567, the line absorbs about 2\% of the continuum and yields a nickel abundance of -6.75 and -5.27.
Nickel appears to be underabundant in the three Chemically Peculiar star but slightly overabundant in the superficially normal HD 174567.

\subsubsection{Gallium}

\begin{figure}[!t]
\epsscale{.80}
\plotone{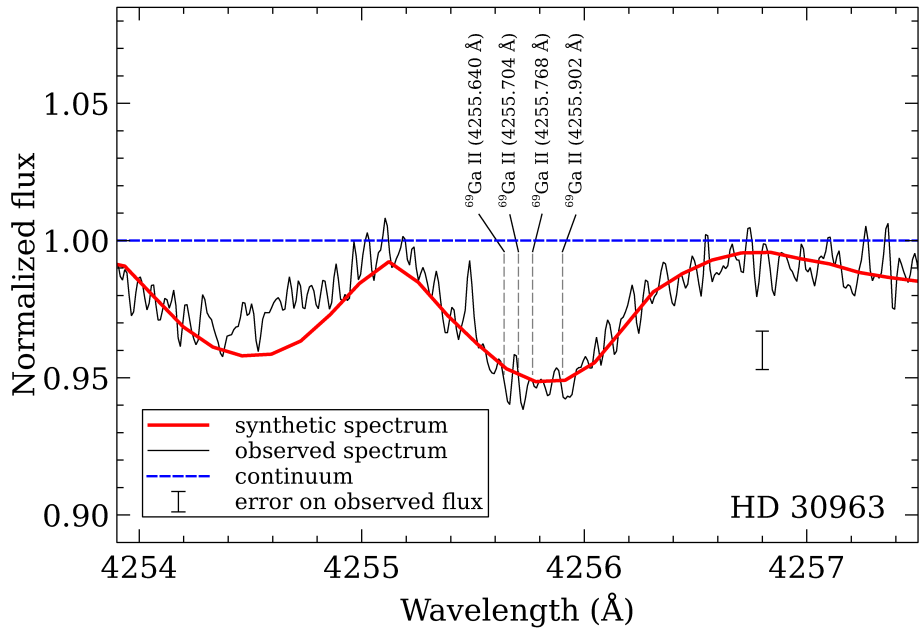}
\caption{Synthesis of the  \ion{Ga}{2} 4255.77  \AA\ blend in HD 30963 showing the contribution of hyperfine structure of gallium.
\label{figure4}}
\end{figure}

\begin{figure}[!t]
\epsscale{.80}
\plotone{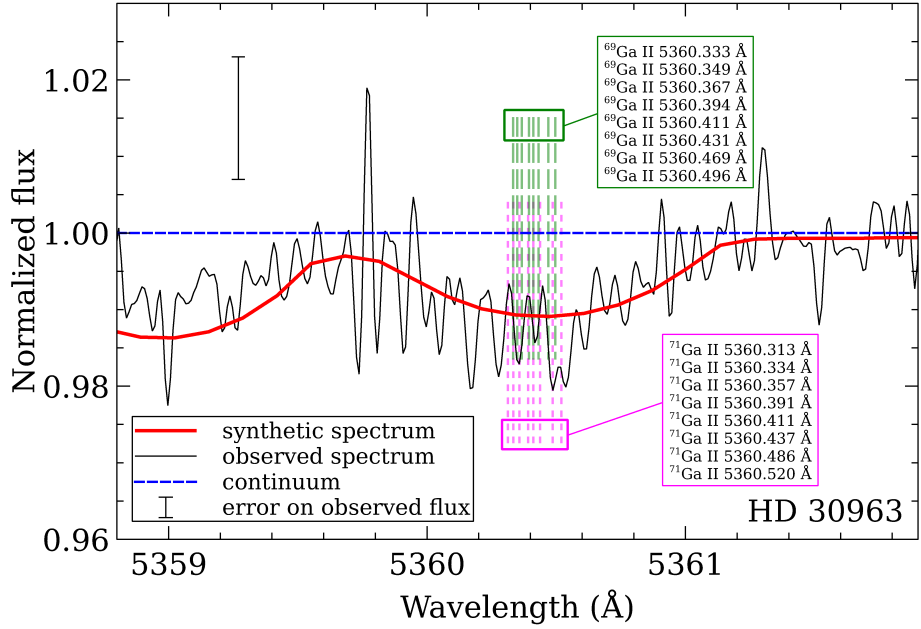}
\caption{Synthesis of the  \ion{Ga}{2} 5360  \AA\ blend in HD 30963 showing the hyperfine structure of gallium.
\label{figure5}}
\end{figure}

We have used the high excitation \ion{Ga}{2} line at $\lambda$ 6334.069 \AA\ with hyperfine splitting of the gallium isotopes. It is the strongest expected line of \ion{Ga}{2} in this temperature regime. The eight transitions retrieved and modified from 
\cite{2000A&A...363..815N} are collected in Table \ref{tab:Gaii}.
\\
In HD 30963, the line of \ion{Ga}{2} at 6334.069 \AA\ is clearly present. Other lines of \ion{Ga}{2} are present at 4251.108 \AA, 4254.032 \AA, 4255.64 \AA\ and 4261.995 \AA\ and 5360.313 \AA, 5363.353 \AA\ and 5421.122 \AA\ in the red part of the spectrum. The fits to the three ines of \ion{Ga}{2} and their hyperfine structures at 4255.6 \AA\ and  5360.3 \AA\ are collected in Figure \,\ref{figure4} and Figure \,\ref{figure5} and consistently yield a large overabundance for HD 30963. In the other three stars, gallium is solar or underabundant.

 % should we give the linelist for this line ?
\begin{table}
\centering
\caption{The \ion{Ga}{2} $\lambda$ 6334.069 \AA\ linelist}
\begin{tabular}{ccccc}
\hline
Wavelength & 	ion 	   & $\log gf$ & lower energy level & EW \\
(\AA) & & & (cm$^{-1}$) & (m\AA) \\ \hline
  6333.930 &  \ion{Ga}{2}  &  0.08 & 102944.595 &   14.7 \\
  6333.980 &  \ion{Ga}{2}  &  0.21 & 102944.595 & 20.5  \\
  6333.990 &  \ion{Ga}{2}  &  0.08 & 102944.595 & 14.6  \\
  6334.069 & \ion{Ga}{2}   &  0.10 & 102944.595 & 15.3 \\
  6334.080 &  \ion{Ga}{2}  &  0.40 & 102944.595 & 26.5  \\
  6334.083 &  \ion{Ti}{2}  & -2.08 &  66521.008 & 0.0   \\
  6334.120 &  \ion{Ga}{2}  &  0.02 & 102944.595 & 12.8 \\
  6334.200 &  \ion{Ga}{2}  &  0.09 & 102944.595 & 20.9  \\
  6334.208 &  \ion{Cr}{2}  & -1.66 &  89812.422 & 0.0  \\
  6334.290 &  \ion{Ga}{2}  &  0.01 & 102944.595 & 16.8  \\
  6334.368 &  \ion{Ni}{2}  & -2.36 & 120271.975 & 0.0  \\ \hline
\label{tab:Gaii}			      		     	  
\end{tabular}
\end{table} 

% Figure for 3 unblended lines of Ga II

% Mercury synthesis figure here 

\subsection{The Sr-Y-Zr triad}

\subsubsection{Strontium}

 The two lines of Multiplet 2 have critical evaluation (NIST quality A and AA).
The abundance of strontium was derived from the fit to the only unblended line of \ion{Sr}{2} at 4215.52 \AA. Strontium is significantly overabundant in HD 30085 and $\chi$ Lupi A and overabundant in HD 30963 and HD 174567. 

\subsubsection{Yttrium}

The yttrium lines are conspicuous in the spectra of HD 30085 and HD 30963, most of them absorbing from a few \% up to 10 \%.
The only unblended lines are the low excitation lines at 3982.592 \AA\ and 5662.925 \AA. Yttrium is very overabundant in each star.

\subsubsection{Zirconium}

The  only two \ion{Zr}{2} unblended lines are $\lambda$ 4443.008 \AA\ and $\lambda$ 4457.431 \AA. Zirconium is overabundant in all stars.

\subsection{Barium}

The two resonance lines of \ion{Ba}{2} (Multiplet 1) at 4554.029 \AA\ and 4934.076 \AA\ and the low excitation line 6141.59 \AA\ are present. The hyperfine structure of the various isotopes of barium we have used is collected in Table \ref{tab:Baii}. Barium is overabundant by a factor of 10 in HD 30085, HD 30963 and HD 174567 and a factor of 5 in $\chi$ Lupi A.

\begin{table}
\centering
\caption{The \ion{Ba}{2} linelists}
\begin{tabular}{ccccc}
\hline
Wavelength & 	ion 	   & $\log gf$ & lower energy level & EW \\ 
(\AA) & & & (cm$^{-1}$) & (m\AA) \\ \hline
   4553.934 &  \ion{Zr}{2} &  -0.57 &  19514.840 & 0.1  \\
   4553.995 &  \ion{Ba}{2} &  -1.57 &      0.000 & 0.1 \\
  4553.997  & \ion{Ba}{2}  &  -1.57 &      0.000 &  0.1 \\
  4553.998  & \ion{Ba}{2}  & -1.99  &     0.000  & 0.0  \\
  4553.999  & \ion{Ba}{2}  & -1.82  &     0.000  & 0.1  \\
  4554.001  & \ion{Ba}{2}  & -1.82  &     0.000  &  0.1 \\
  4554.001  & \ion{Ba}{2}  &  -2.22 &      0.000 & 0.0  \\
  4554.011  & \ion{Cr}{1}  &  -0.73 &  33040.093 & 0.0   \\
  4554.029  & \ion{Ba}{2}  &  0.02  &     0.000  & 16.6  \\
  4554.029  & \ion{Ba}{2}  & -1.45  &     0.000  & 1.3  \\
  4554.029  & \ion{Ba}{2}  & -0.94  &     0.000  & 3.8   \\
  4554.046  & \ion{Ba}{2}  & -1.38  &     0.000  & 0.3   \\
  4554.049  & \ion{Ba}{2}  & -1.82  &     0.000  & 0.1  \\
  4554.049  & \ion{Ba}{2}  & -1.15  &     0.000  & 0.6  \\
  4554.050  & \ion{Ba}{2}  & -2.52  &     0.000  & 0.0  \\
  4554.051  & \ion{Ba}{2}  & -1.57  &     0.000  & 0.1  \\
  4554.052  & \ion{Ba}{2}  & -2.29  &     0.000  & 0.0  \\
\hline 
  4934.005  & \ion{Fe}{1}  & -0.61  & 33507.120  &  0.4 \\
  4934.074  & \ion{Ba}{2}  & -3.12  &     0.000  &    0.0 \\
  4934.074  & \ion{Ba}{2}  & -1.33  &     0.000  & 0.8  \\
  4934.075  & \ion{Ba}{2}  & -3.15  &     0.000  & 0.0 \\
  4934.075  & \ion{Ba}{2}  & -1.10  &     0.000  & 1.3  \\
  4934.076  & \ion{Ba}{2}  & -1.77  &     0.000  & 0.3  \\
  4934.076 &  \ion{Ba}{2}  & -1.25  &     0.000  &  0.9  \\
  4934.077 &  \ion{Ba}{2}  & -0.29  &     0.000  &  7.0  \\
  4934.084  & \ion{Fe}{1}  &  -2.30 &  26627.608 &   0.1 \\
\hline
  6141.597  &  \ion{Fe}{1} &   -3.12 &  33765.304 & 0.0  \\
  6141.713  & \ion{Ba}{2}  & -0.22  &  5674.824   &    2.4 \\
  6141.713  & \ion{Ba}{2}  & -1.03  &  5674.824   &     0.4 \\
  6141.714  & \ion{Ba}{2}  & -1.18  &  5674.824   &    0.3   \\
  6141.714  & \ion{Ba}{2}  & -1.26  &  5674.824  & 0.2   \\
  6141.715  & \ion{Ba}{2}  & -1.69  &  5674.824  &  0.1  \\
  6141.717  & \ion{Ba}{2}  & -3.07  &  5674.824  & 0.0  \\
  6141.718  & \ion{Ba}{2}  & -3.05  &  5674.824  & 0.0 \\
  6141.731  & \ion{Fe}{1}  &  -1.61 & 29056.322  & 0.1 \\ \hline
\label{tab:Baii}			      		     	  
\end{tabular}
\end{table}

\subsection{The Rare Earths}

We have searched for once ionized rare earths elements in the spectra of the four stars, namely \ion{La}{2}, \ion{Ce}{2}, \ion{Pr}{2}, \ion{Nd}{2}, \ion{Sm}{2}, \ion{Eu}{2}, \ion{Gd}{2}, \ion{Tb}{2}, \ion{Dy}{2}, \ion{Ho}{2} and \ion{Er}{2}. As the twice ionised rare earths often are the dominant stages in these atmospheres of late B stars, we also looked for the twice ionized species, \ion{Pr}{3} and \ion{Nd}{3}, using the lines listed either in NIST or in DREAM and the linelists published in \cite{RRKB} and \cite{2007A&A...473..907R} .

\subsubsection{Lanthanum}

The spectrum synthesis of the \ion{La}{2} line at 4042.91 \AA\ feature is consistent with one solar abundance upper limit for HD 30085 and an overabundance. Lanthanum is overabundant in HD 174567 and $\chi$ Lupi A. The line is blended in HD 30963.
 \\
For \ion{La}{3}, the three lines $\lambda$ 4499.064 \AA, $\lambda$ 5145.718 \AA\ and $\lambda$ 5148.391 \AA\ retrieved from NIST are all blended with more abundant species in the three stars.

\subsubsection{Cerium}

The synthesis of the unblended  \ion{Ce}{2} lines at 4133.80 \AA\ and 4460.21 \AA\ yields large overabundances of cerium  in HD 30085 and $\chi$ Lupi A and a one solar upper limit for HD 30963 and HD 174567.
\\
For \ion{Ce}{3}, we find coincidences for seven out of ten low excitation lines listed in DREAM, however all of them are heavily blended in the spectra of the four stars.

\subsubsection{Praseodymium}

The lines of \ion{Pr}{3} at 5284.69 and 5299.99 \AA\ are unblended and provide large overabundances in HD 30085, HD 174567 and $\chi$ Lupi A. They are consistent with a solar abundance upper value for HD 30963.

\subsubsection{Neodymium}

For \ion{Nd}{3}, the strongest expected lines in DREAM are the resonance lines $\lambda$ 5265.019 \AA\ and $\lambda$ 5294.099 \AA\ which correspond to structures absorbing respectively 3\% and 5\% of the local continuum inside complex blends. The synthesis of these lines yields moderate to large overabundances of \ion{Nd}{3} in HD 30085, in HD 30963, HD 174567 and in $\chi$ Lupi A.

\subsubsection{Samarium}

The \ion{Sm}{2} lines at 4280.79 \AA\ and 4424.32 \AA\ are unblended  and are consistent with a 1 solar upper limit abundance of samarium in HD 30085, HD 30963, HD 174567 and an overabundance of 25 in $\chi$ Lupi A.

\subsubsection{Europium}

The \ion{Eu}{2} resonance line at 4129.73 \AA\ is unblended and consistent with a 1 solar upper limit abundance of europium in HD 30085, HD 30963, HD 174567 and a significant overabundance of about 100 in $\chi$ Lupi A.

\subsubsection{Gadolynium}

The \ion{Gd}{2} line at 4037.32 \AA\ is unblended and properly reproduced with large overabundances of about 100 in HD 30085, HD 174567 and $\chi$ Lupi A. It is blended in HD 30963.

\subsubsection{Terbium}

The \ion{Tb}{2} line at 4005.47 \AA\ is unblended in HD 30085 and HD 30963 and provides a 1 solar upper limit abundance. It is blended in HD 174567 and $\chi$ Lupi A.
\\

\subsubsection{Dysprosium}

The \ion{Dy}{2} line at 4000.45 \AA\ is unblended in HD 30085, HD 174567 and $\chi$ Lupi A and provides a 1 solar upper linit for HD 30085 and an overabundance by a factor of 60 respectively for HD 174567 and $\chi$ Lupi A. This line is blended in HD 30963.

\subsubsection{Holmium}

The low excitation line of \ion{Ho}{2} at 4152.62 \AA\ is unblended and provides a 1 solar upper limit in HD 30085, HD 30963 and HD 174567 and an overabundance in $\chi$ Lupi A.
\\
%For \ion{Ho}{3} all the resonance lines listed in DREAM and are also consistent with a 1 solar upper limit in these stars.

\subsubsection{Erbium}

For \ion{Er}{2} ,  the line at 4142.91 \AA\ is consistent with a 1 solar upper limit for HD 30085, is blended in HD 30963 and $\chi$ Lupi A and reproduced by a large overabundance of about 173 in HD 174567. 

\subsubsection{Thulium}

The line at 4242.15 \AA\ of \ion{Tm}{2} is blended in HD 30085 and $\chi$ Lupi A and consistent with a 1 solar upper limit for HD 30963 and HD 174567.
 
\subsubsection{Ytterbium}
 
 The \ion{Yb}{2} line at 4135.095 \AA\ provides a one solar upper limit for HD 30085, HD 174567 and $\chi$ Lupi A. It is blended in HD 30963. 

\subsubsection{Hafnium}

The \ion{Hf}{2} line at 3918.08 \AA\ yields a 1 solar upper limit for HD 30085 and HD 174567 and is blended in HD 30963 and $\chi$ Lupi A.

\subsection{The very heavy elements Osmium, Platinum, Gold and Mercury}

\subsubsection{Osmium}

The \ion{Os}{2} line at 4158.44 \AA\ is consistent with a 1 solar upper limit in HD 30085, HD 174567 and $\chi$ Lupi A and is blended in HD 30963.

\subsubsection{Platinum}
 
%%\cite{1969ApJ...156L.101D} analysis of HR 4072 and  \cite{1973ApJ...181..811D} linelist of \ion{Pt}{2} for HR 4072 A and $\chi$ Lupi A A. We also used 
%astrophysical absolute oscillator strengths for optical \ion{Pt}{2} lines 
%published in \cite{1984PhST....8...39D}
%and
%Engelman's (1989) measurements of the hyperfine structure and isotope splittings of 17 \ion{Pt}{2} lines. \cite{2000ApJ...539..908W} also provide in their table 5 hyperfine structure for 3 \ion{Pt}{2} %lines with oscillator strengths retrieved from \cite{1984PhST....8...39D}. This allowed the compilation of a list of 8 lines, a few of which have hyperfine structure and isotopic shifts evaluated.
%Seven lines of \ion{Pt}{2} could be identified in the spectra of HD 30085, the strongest one being $\lambda$ 4514.47 \AA\ is unblended and coincides with a well defined feature absorbing about 2.5 \% %of the local continuum which we have used to estimate the platinum overabundance. The other lines are all blended. 

The 4514.47 \AA\ \ion{Pt}{2} line is the only unblended line of platinum in the spectra of the four stars. It does not have any hyperfine structure published for the various isotopes of Pt. We therefore derived a crude estimate of the platinum overabundance. There are no other species at this wavelength which could account for the observed absorption as shows the spectrum synthesis without the \ion{Pt}{2} line. The wavelength scale around the \ion{Pt}{2} line was checked using the \ion{Fe}{2} neighbouring control lines whose NIST wavelengths are accurate to within $\pm$ 0.001 \AA.

\begin{figure}
\epsscale{.80}
\plotone{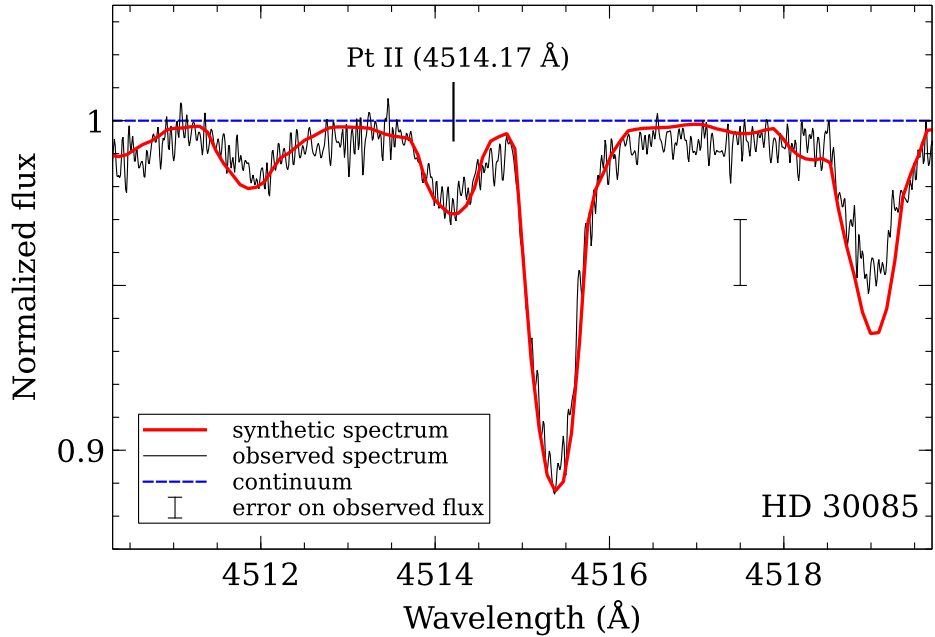}
\caption{Synthesis of the \ion{Pt}{2} 4514.17 \AA\ line in HD 30085. 
\label{figure6}}
\end{figure}

The \ion{Pt}{2} line at 4046.45 \AA\ has hyperfine structure published (5 transitions), but it is blended with the 9 hyperfine transitions of the \ion{Hg}{1} line at 4046.57 \AA. Using the mercury abundance obtained in the following paragraph, the synthesis of this blend in the spectra of HD 30085 and HD 30963 also yields an overabundance similar to that derived from the 4514.47 \AA\ line. 
The final adopted platinum abundance is that derived from the \ion{Pt}{2} line at 4514.47 \AA. It ranges from large overabundances of the order of $10^4$ for 30085 and $10^3$ in HD 30963 to a solar upper limit (-10.65) for HD 174567. This line is blended in $\chi$ Lupi A. The synthesis of the \ion{Pt}{2} line at 4514.47 \AA\ in HD 30085 is shown in Figure \,\ref{figure6}. 

% Platinum figure here 

\subsubsection{Gold}

The only \ion{Au}{2} line corresponding with weak features is $\lambda$ 4052.790 \AA\ but is blended with an \ion{Fe}{1} line in the spectra of all four stars.
 
% REPRENDRE ICI

\subsubsection{Mercury}

The abundance of mercury has been derived from the low excitation \ion{Hg}{2} line at 3983.93 \AA\ only. This feature absorbs about 12 \% of the continuum in HD 30085, HD 30963 and $\chi$ Lupi A and about 4 \% in HD 174567. The other \ion{Hg}{2} lines are all high excitation lines either weak or blended with more abundant species and were not synthesized. In particular \ion{Hg}{2} 6149.4749 \AA\ is blended with \ion{Fe}{2} 6149.258 \AA\ in HD 30085, HD 30963 and HD 174567 which precludes any conclusion on the mercury abundance. The blend is resolved in $\chi$ Lupi A and yields an overabundance which agrees from that found with the 3983.93 \AA\ line.
\\
To model the \ion{Hg}{2} line at 3983.93 \AA, we have included 9 transitions ie., all hyperfine structure from the various isotopes from $\mathrm{Hg}^{196}$  and $\mathrm{Hg}^{204}$ 
listed in \cite{Do03}. These transitions are collected together with blending lines from \ion{Ti}{1}, \ion{Fe}{1}, \ion{Cr}{1} and \ion{Cr}{2} in Table \ref{tab:Hgii}.
% insert Hg II linelist here

\begin{table}
\centering
\caption{The \ion{Hg}{2} 3983.93 \AA\ linelist}
\begin{tabular}{ccccc}
\hline
Wavelength & 	ion 	   & $\log gf$ & lower energy level & EW \\
(\AA) & & & (cm$^{-1}$) & (m\AA) \\ \hline
  3983.771 &  \ion{Hg}{2}  &  -4.50   &    35514.999 &   1.2 \\ 
  3983.827 &   \ion{Ti}{1} & 	-1.46 &   17075.258  &    0.0 \\ 
  3983.832 &   \ion{Fe}{1} & 	-4.81 &   24338.766  &    0.0  \\ 
  3983.839 &   \ion{Hg}{2} &   -3.00  &    35514.000 &   14.1 \\ 
  3983.844 &   \ion{Hg}{2} &   -3.13  &    35514.000 &   12.5 \\ 
  3983.851 &   \ion{Cr}{2} &   -4.31  &    81962.291 &   0.0 \\ 
  3983.853 &   \ion{Hg}{2} &   -3.00  &    35514.000 &   14.1 \\ 
  3983.874 &   \ion{Fe}{1} & 	-2.70 &   34039.515  &  0.0  \\ 
  3983.899 &   \ion{Cr}{1} & 	 0.35 &   20520.904  &  3.2 \\ 
  3983.912 &   \ion{Hg}{2} &   -2.50  &  35514.000   &  19.2 \\ 
  3983.932 &   \ion{Hg}{2} &   -3.10  &  35514.000   & 12.9  \\ 
  3983.941 &   \ion{Hg}{2} &   -2.90  &  35514.000   & 15.3  \\ 
  3983.956 &   \ion{Fe}{1} & 	-1.02 &   21999.129  &  2.5  \\ 
  3983.986 &   \ion{Cr}{2} &   -2.32  &  90512.561   &   0.0  \\ 
  3983.991 &   \ion{Cr}{2} &   -2.88  &  82362.188   &    0.0   \\ 
  3983.993 &   \ion{Hg}{2} &   -3.00  &  35514.000   & 14.2  \\ 
  3984.022 &   \ion{Cr}{1} & 	-2.47 &   35397.971  &  0.0  \\ 
  3984.072 &   \ion{Hg}{2} &   -3.00  &  35514.000   & 14.4  \\ \hline
\label{tab:Hgii}			      		     	  
\end{tabular}
\end{table}

The wavelength scale was checked by using four control lines on each side of the \ion{Hg}{2} line: shortwards the \ion{Zr}{2} line at 3982.0250 \AA\ , the \ion{Y}{2} line at 3982.59 \AA\ and longwards the \ion{Zr}{2} lines at 3984.718 \AA\ and 3991.15 \AA. Once corrected for the radial velocities of HD 30085 and HD 30963, the centers of these lines are found at their expected laboratory locations to within $\pm$ 0.02 \AA\  which we will adopt as the accuracy of the wavelength scale in this spectral region.
After rectification to the red wing of the $H_{\epsilon}$ line, the core of the \ion{Hg}{2} lines of HF 30085 and HD 30963 are flat and extend from about 3983.90 $\pm$ 0.02 \AA\  to 3984.07 $\pm$ 0.02 \AA\ which roughly correspond to the positions of lines of the heaviest isotopes $\mathrm{Hg}^{200}$  and $\mathrm{Hg}^{204}$ . 
The heavy isotopes of Hg thus contribute more to the absorption than the lighter isotopes in HD 30085 and HD 30963 as is the case in the coolest HgMn stars (\citealt{1999ApJ...521..414W}, \citealt{1976ApJ...204..131W}). In $\chi$ Lupi A, the \ion{Hg}{2} line center is at 3984.07 \AA\ which corresponds to the line of the heaviest isotope.
\\
First, we checked the influence of possible contaminant species to the \ion{Hg}{2} line from 3983.50 \AA\ up to 3984.50 \AA\ which we estimate to be the maximum extension of the line wings. The only possible contaminants are three weak lines: \ion{Fe}{2}  $\lambda$  3983.737 \AA, \ion{Cr}{1} $\lambda$ 3983.897 \AA\  and \ion{Fe}{1} $\lambda$ 3983.956 \AA. The equivalent widths of these lines were computed for the derived Fe and Cr abundances and the sum of their contributions, which amounts to about 4.9 m\AA, is by far insufficient to reproduced the equivalent width of the observed feature at 3983.93 \AA\ (about 64 to 70  m\AA) of the \ion{Hg}{2} in HD 30085, HD 30963 and $\chi$ Lupi A.
Another test consisted in computing the flux without the \ion{Hg}{2} transitions. This consistently resulted into a very weak absorption feature from 3983.50 \AA\ to 3984.50 \AA\ in the four stars. We can therefore conclude that the observed features at 3983.93 \AA\ is indeed mostly due to the \ion{Hg}{2} line and is almost free of blending.
\\
Including the hyperfine structure of the seven isotopes from Table 2 of \cite{Do03} significantly reduces the mercury abundance, for example to about 120000 $\odot$ in HD 30085 (it would be of the order of $10^{7}$ $\odot$ if we include the single transition retrieved from VALD). In HD 30085 and HD 30963, the 12 \% absorption at the core is reproduced but the synthetic profile is significantly offset to the blue by about -0.05 \AA\ from the observed blue wing and + 0.10 \AA\ from the observed red wing. In order to shift redwards the synthetic line core and give it a flat shape comparable to the observed one, we have iteratively altered the oscillator strengths of the individual hyperfine components until the overall observed line profile could be reproduced. We progressively decreased the oscillator strengths of the first 4 transitions of the three lightest isotopes $\mathrm{Hg}^{196}$, $\mathrm{Hg}^{198}$ and $\mathrm{Hg}^{199}$
and increased those of 5 transitions of the heaviest 
$\mathrm{Hg}^{200}$ , $\mathrm{Hg}^{201}$, $\mathrm{Hg}^{202}$  and $\mathrm{Hg}^{204}$ in order to redistribute the computed line opacity towards longer wavelengths. The final combination of oscillator strengths is recorded in Table \ref{tab:Hgii}.

\begin{figure}[!t]
\epsscale{.80}
\plotone{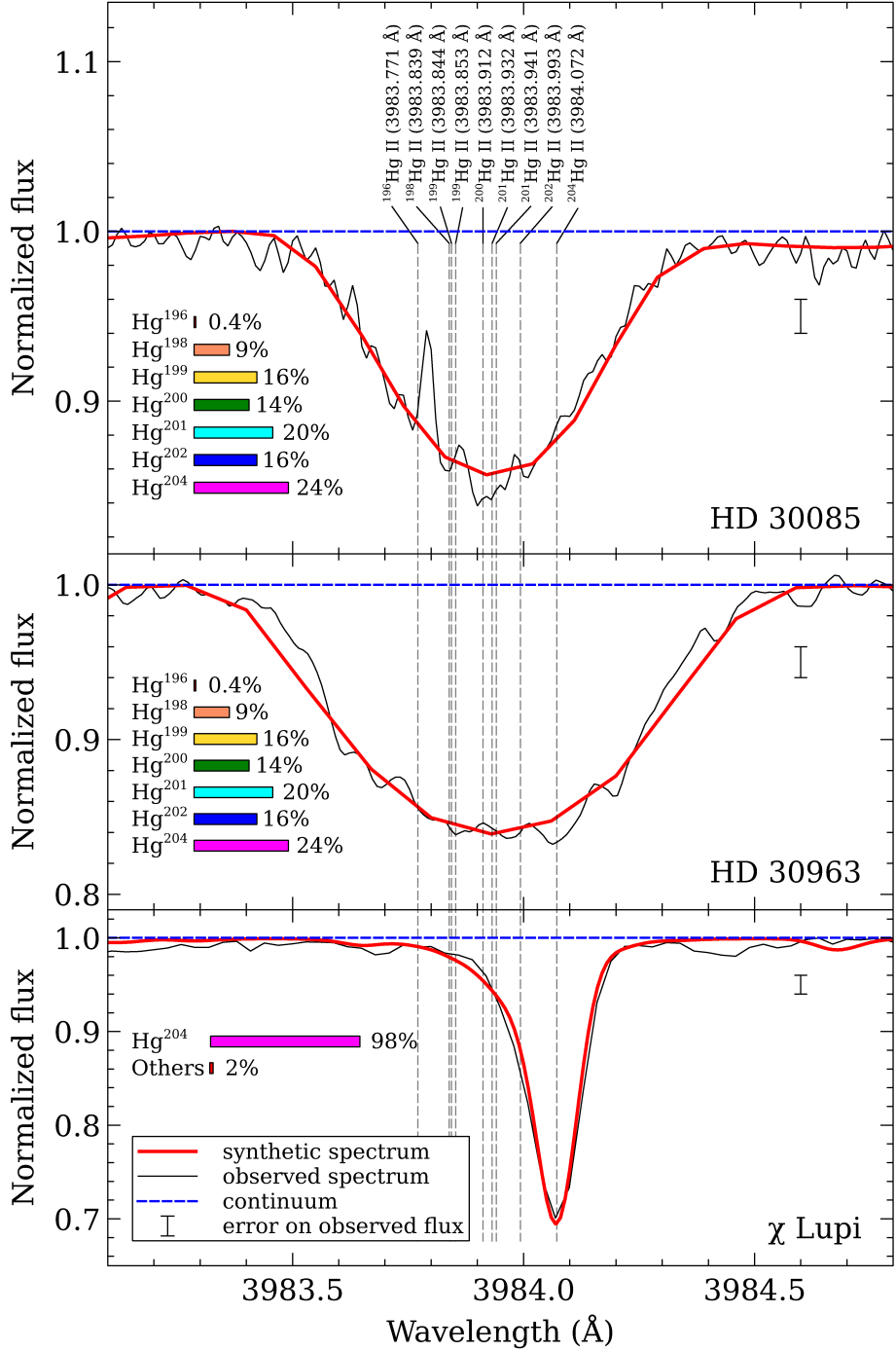}
\caption{Synthesis of the  \ion{Hg}{2}3983.87 \AA\ blend in HD 30085, HD 30963 and $\chi$ Lupi A A showing the contribution of hyperfine structure of various isotopes.
\label{figure7}}
\end{figure}

\begin{figure}[!t]
\epsscale{.80}
\plotone{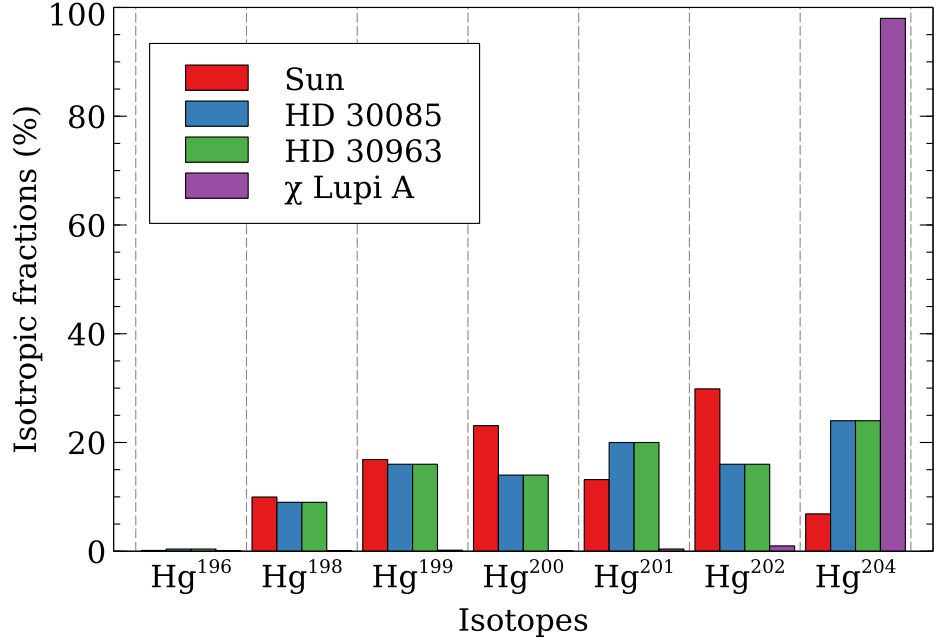}
\caption{Possible contribution of the various isotopes of Hg to the 3983.93 \ion{Hg}{2} line in each star.
\label{figure8}}
\end{figure}

We can give a very rough estimate of the contribution of each isotope in HD 30085 and HD 30963 by dividing the equivalent widths of its hyperfine components to the total equivalent width of the whole line. We thus find increasingly larger contributions as we move to heavier isotopes: 
$\mathrm{Hg}^{196}$ 0.4 \%, $\mathrm{Hg}^{198}$ 9 \%, $\mathrm{Hg}^{199}$ 16 \%, $\mathrm{Hg}^{200}$ 14 \%, $\mathrm{Hg}^{201}$ 20 \%, $\mathrm{Hg}^{202}$ 16 \% and $\mathrm{Hg}^{204}$ 24 \%.
\\
The final mercury abundances derived from the synthesis of the 3983.93 \AA\ 
yield very large overabundances of about -5.83 in HD 30085 (see Figure \,\ref{figure7}), -5.31 in HD 30963, -5.91 for $\chi$ Lupi A and -7.61 in HD 174567. The observed \ion{Hg}{2} line center in $\chi$ Lupi A is at about 3984.072 \AA, which corresponds to the heaviest isotope of mercury only. The possible contribution of the various isotopes of mercury in HD 30085, HD 30963 and $\chi$ Lupi A is represented as histograms in Figure \,\ref{figure8}.

% Mercury synthesis figure here 

\section{Discussion}

In table \ref{tab:compChiLupi} we compare the abundances we derived for $\chi$ Lupi A with those derived from high resolution spectra in the optical range by \cite{wahlgren1994}.

\begin{deluxetable}{l|c|c||l|c|c}
\tabletypesize{\normalsize}
\tablecaption{Comparison between the abundances of $\chi$ Lupi A and the ones derived in \cite{wahlgren1994}}
\tablecolumns{2}\tablewidth{12cm}
\tablehead{Element&\multicolumn{2}{c||}{$<\log(X/H)>$}&Element&\multicolumn{2}{c}{$<\log(X/H)>$}\\
& This work & Wahlgren et al. & & This work & Wahlgren et al. \\}
\startdata
\ion{He}{1}& -1.67$\pm$0.23&\ion{C}{2}&-3.76$\pm$0.34&-3.85\\
\ion{O}{1}&-3.53$\pm$0.10&-3.93&\ion{Na}{1}&-4.97&-\\
\ion{Mg}{2}&-4.66$\pm$0.15&-4.60$\pm$0.06&\ion{Al}{2}&-5.79$\pm$0.20&-5.85\\
\ion{Si}{2}&-4.55$\pm$0.05&-4.42$\pm$0.28&\ion{P}{2}&-5.47$\pm$0.16&-\\
\ion{S}{2}&-4.96$\pm$0.08&-4.78$\pm$0.04&\ion{Ca}{2}&-5.83$\pm$0.16&-5.70\\
\ion{Sc}{2}&-10.13$\pm$0.14&-10.16$\pm$0.17&\ion{Ti}{2}&-6.73$\pm$0.06&-6.65$\pm$0.25\\
\ion{V}{2}&-8.02$\pm$0.12&-7.96$\pm$0.23&\ion{Cr}{2}&-6.14$\pm$0.17&-5.96$\pm$0.27\\
\ion{Mn}{2}&-6.13$\pm$0.04&-6.15$\pm$0.13&\ion{Fe}{2}&-4.57$\pm$0.16&-4.29$\pm$0.21\\
\ion{Ni}{2}&-6.52$\pm$0.14&-5.96$\pm$0.10&\ion{Ga}{2}&-9.12&-\\
\ion{Sr}{2}&-7.43$\pm$0.09&-7.03$\pm$0.14&\ion{Y}{2}&-8.06$\pm$0.16&-7.89$\pm$0.16\\
\ion{Zr}{2}&-8.70$\pm$0.09&-8.68$\pm$0.18&\ion{Ba}{2}&-9.17$\pm$0.25&-8.80\\
\ion{La}{2}&-9.13$\pm$0.13&-&\ion{Ce}{2}&-8.90$\pm$0.25&-\\
\ion{Pr}{3}&-9.09$\pm$0.13&-&\ion{Nd}{3}&-8.80$\pm$0.25&-\\
\ion{Sm}{2}&-8.59$\pm$0.13&-&\ion{Eu}{2}&-9.61$\pm$0.25&-\\
\ion{Gd}{2}&-8.70$\pm$0.13&-&\ion{Dy}{2}&-9.08$\pm$0.25&-\\
\ion{Ho}{2}&-9.50$\pm$0.13&-&\ion{Yb}{2}&-10.92$\pm$0.25&-\\
\ion{Os}{2}&-10.55$\pm$0.13&-&\ion{Au}{2}&-10.99$\pm$0.25&-7.49\\
\ion{Hg}{2}&-5.91$\pm$0.13&-6.34&&&\\
\enddata
\label{tab:compChiLupi}
\end{deluxetable}

\begin{figure}
\epsscale{.80}
\plotone{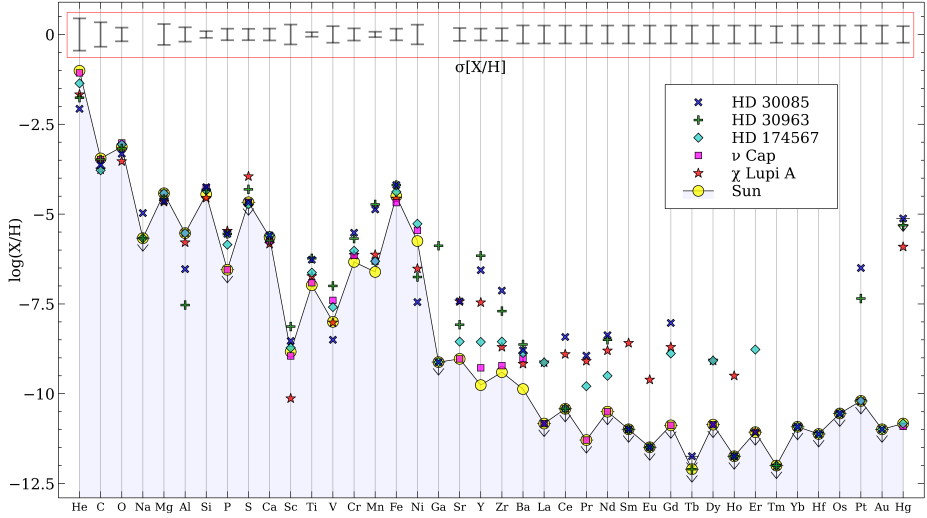}
\caption{Comparison of the abundance patterns of HD 30085, HD 30963, $\chi$ Lupi A, HD 174567 and $\nu$ Cap
\label{figure9}}
\end{figure}

Our abundance determinations for carbon, magnesium, aluminium, silicon, calcium, scandium, titanium, vanadium, chromium, manganese and zirconium  agree within $\pm$ 0.20 dex (a representative error on our abundances) with those derived by \cite{wahlgren1994}. For the other nine elements common to both studies, the abundances differ by more than 0.20 dex. This probably comes from using different lines or different more recent atomic data for lines in common, different model atmospheres and different line synthesis codes.
\\
From the list of identifications provided for the final composition of HD 30085 in table \ref{tab:ident}, we can conclude that the following species are present in the spectra of HD 30085 (and HD 30963 as well): \ion{C}{2}, \ion{N}{2}, \ion{O}{1}, \ion{Na}{1}, \ion{Mg}{2}, \ion{Mg}{1},\ion{Mg}{2}, \ion{Si}{2}, \ion{P}{2}, \ion{S}{2}, \ion{Ca}{2}, \ion{Sc}{2}, \ion{Ti}{2}, \ion{V}{2}, \ion{Cr}{2}, \ion{Cr}{1}, \ion{Mn}{2}, \ion{Fe}{2}, \ion{Ni}{2}, \ion{Ga}{2}, \ion{Sr}{2}, \ion{Y}{2}, \ion{Zr}{2}, \ion{Ba}{2}, \ion{Pt}{2}, \ion{Hg}{2}. Among these species, \ion{Si}{2}, \ion{Ti}{2}, \ion{Cr}{2}, \ion{Mn}{2}, \ion{Fe}{2}, \ion{Y}{2}, \ion{Zr}{2}, have numerous and strong lines and are important opacity sources. Lines from neutrals: \ion{Cr}{1}, \ion{Mn}{1}, \ion{Fe}{1} are also observed and yield lower abundances than \ion{Cr}{2}, \ion{Mn}{2} and \ion{Fe}{2}, indicating possibly departures from LTE. Gallium is present with a large overabundance in HD 30963.
\\
In HD 30085, we only find evidence for overbundances of four Rare Earths from unblended lines of \ion{Ce}{2}, \ion{Pr}{3} and \ion{Nd}{3} and \ion{Gd}{2}. For the other Rare Earths, we find upper limits: the abundances must be solar or lower. In HD 30963, the blending is higher because of the larger $\vsini$, we can conclude only that \ion{Nd}{3} may be present and overabundant (evidence from one line only), all other lines are blended. In HD 174567, the low $\vsini$ favors detection of weak unblended lines and evidence is found for overabundances of \ion{La}{2}, \ion{Pr}{3}, \ion{Nd}{3}, \ion{Gd}{2}, \ion{Dy}{2} and \ion{Er}{2}. 
The presence of once ionised Rare Earths is difficult to assess at these temperatures above 10000 K
because they are not the dominant ionisation stage. In general, the abundances for the Rare Earths should be regarded as the least reliable ones in this study because they were usually infered from the synthesis of one weak line of the once ionised specie only.
\\
The overabundances found for HD 30085 and HD 30963 run from very mild (2.0 $\odot$  for Fe) up to quite large 
($10^{5}$ $\odot$ for Hg). The underabundances run from mild -0.80 $\odot$ for carbon to pronounced -0.02 $\odot$ for nickel. Helium is quite underabundant in HD 30085, HD 30963 and $\chi$ Lupi A. 
\\
The abundance patterns of the five stars are compared in Figure \,\ref{figure9}. The comparison bona-fide normal late B star, $\nu$ Cap, has a nearly solar composition for all elements as already found by  \cite{2018ApJ...854...50M} and \cite{1991MNRAS.252..116A}. For the other four stars, the overall trend is that the light elements (He through Mg) are most often defficient whereas elements heavier than Mn tend to be more and more overabundant as the atomic mass increases (with the exception of nickel).
This pattern is most likely caused by radiative diffusion.
In particular the  overall shapes of the patterns of HD 30085 and HD 30963 compare well with that of $\chi$ Lupi A  (HD 141556), even though differences for individual abundances exist.
%HD 141556 is a well-studied HgMn star whose effective temperature and surface gravity are close to those of HD 30085 and HD 30963. 
There are several similarities for the abundances of the elements lighter than Scandium between $\chi$ Lupi A and HD 30963, in particular for helium, magnesium, phosphorus, calcium. For elements heavier than scandium, the abundances of HD 30963 differ from that of $\chi$ Lupi A. For HD 30085, the carbon, magnesium, phosphorus, gallium, strontium, praseodymium abundances are similar to those of $\chi$ Lupi A. Overall, the chemical patterns of HD 30085 and HD 30963 follow that of $\chi$ Lupi A, which definitely confirms that these two star are new HgMn stars. Moreover the patterns of HD 30085 and HD 30963 also fall well within the general abundance pattern of confirmed HgMn stars in Fig.~1 of \cite{Ghazaryan2016} compilation. HD 30085 and HD 30963 are most likely classical HgMn stars.

 The pattern of HD 174567 is intermediate between that of $\nu$ Cap and that of the three HgMn stars. We therefore propose that HD 174567 is a new mild Chemically Peculiar star with overabundances of elements heavier than strontium less pronounced than in HD 30085 and HD 30963 but definitely larger than in $\nu$ Cap. The overabundances for Sr, Y, Zr, several Rare Earths and mercury suggest that HD 174567 could be a new cool and mild HgMn star. More observations of HD 174567 are foreseen in order to elucidate the nature of this interesting object.

%describe the abundance pattern found, compare to another HgMn of similar Teff, %logg (chi Lupi?)

\section{Conclusion}

With their characteristic underabundances of most light elements up to calcium and overabundances of manganese and elements heavier than strontium, in particular platinum and mercury, HD 30085 and HD 30963 are definitely two new HgMn stars. Their abundances differ from those of $\chi$ Lupi A possibly because their initial abundances differed slightly and because radiative diffusion operated in a slightly different manner in the younger and more massive stars HD 30085 and HD 30963. For HD 174567, the mild deficiencies for light elements and overabundances for strontium, yttrium and zirconium and several rare-earths and mercury suggest that this object should be reclassified as a mild Chemically Peculiar star. More observations of this interesting star will help elucidate its nature.

\label{sec:conclusion}

%\appendix
\begin{appendix}
\section{Determination of uncertainties}
For a representative line of a given element, six major sources are included in the uncertainty determinations: the uncertainty on the effective temperature ($\sigma_{T_{\rm{eff}}}$), on the surface gravity ($\sigma_{\log g}$), on the microturbulent velocity ($\sigma_{\xi_{t}}$), on the apparent rotational velocity ($\sigma_{v_{e}\sin i}$), the oscillator strength ($\sigma_{\log gf}$) and the continuum placement ($\sigma_{cont}$). These uncertainties are supposed to be independent, so that the total uncertainty $\sigma_{tot_{i}}$ for a given transition (i) verifies:\\
\begin{equation}
\sigma_{tot_{i}}^{2}=\sigma_{T_{\rm{eff}}}^{2}+\sigma_{\log g}^{2}+\sigma_{\xi_{t}}^{2}+\sigma_{v_{e}\sin i}^{2}+\sigma_{\log gf}^{2}+\sigma_{cont}^{2}.
\end{equation} 
The mean abundance $<[\frac{X}{H}]>$ is then computed as a weighted mean of the individual abundances [X/H]$_{i}$ derived for each transition (i):\\
\begin{equation}
<[\frac{X}{H}]>=\frac{\sum_{i}([\frac{X}{H}]_{i}/\sigma_{tot_{i}}^{2})}{\sum_{i}(1/\sigma^{2}_{tot_{i}})}
\end{equation}
and the total error, $\sigma$ is given by: \\
\begin{equation}
\frac{1}{\sigma^{2}}=\sum_{i=1}^{N}(1/\sigma_{tot_{i}}^{2})  
\end{equation}
where N is the number of lines per element. The uncertainties $\sigma$ for each element are collected in Tab.~\ref{uncer}.\\
\end{appendix}

% temporarily place Table uncertainties here 

% new corrected table here:

\begin{acknowledgements} 
RM thanks Pr. Charles Cowley for his insightful comments during the analysis of HD 30085 and HD 174567.
We thank the OHP night assistants for their helpful support during the three observing runs.
This work has made use of the VALD database, operated at Uppsala University, the Institute of Astronomy RAS in Moscow, and the University of Vienna.
We have also used the NIST Atomic Spectra Database (version 5.4) available http://physics.nist.gov/asd.
We also acknowledge the use of the ELODIE archive at OHP available at http://atlas.obs-hp.fr/elodie/.
\end{acknowledgements}

\bibliography{ref.bib}
\bibliographystyle{aa}

\clearpage 

% Figures here before tables

% All figures here in increasing order
\newpage

\clearpage
% Tables

%-----------------------------------------------------------------------------------------------------------------------------------------------------------------

% new uncertainty table here

% temporarily place Table uncertainties here 

% new corrected table here:

\setcounter{table}{6}
\begin{table}
\centering
\caption{Abundance uncertainties for the elements analysed in  HD 30085, HD 30963, HD 174567 and HD 141556}
\begin{tabular}{ccccccccccc}

%\tabletypesize{\normalsize}   
%\tablehead{}

%\tablecolumns{10}\tablewidth{12cm}  
%\tablehead{}
%\startdata
\hline
  && HeI & CII&OI&MgII&AlII&SiII &PII &SII& CaII \\ \hline
$\Delta$ \teff  & +200 K & -0.30  & 0.00  & 0.05    & 0.047 &0.075  & -0.017 & 0.00 &  0.00 &  0.109\\
$\Delta \log g$ &  0.15 dex& 0.079  & 0.24  & -0.06   & -0.052   & 0.118  & -0.017 & -0.12  & -0.12 & -0.032\\
$\Delta \xi_{t}$ & +0.20 \kms & 0.00 & 0.00  & 0.00   & 0.00 &  0.075  & -0.035  & -0.05 & -0.046 & 0.00 \\
$\Delta \log gf $ & +0.10     & -0.30 & -0.12  & -0.16  & -0.25 & -0.028   & -0.053 & -0.07 &  -0.071 & -0.105  \\
$\Delta continuum$ &          & -0.12 & 0.21  & 0.075    & 0.13 & 0.25  & 0.062 & 0.06  & 0.061 & 0.058  \\
\hline      
$2\sigma_{[X/H]}$  & & 0.45  & 0.34 &  0.19 & 0.29  & 0.20 & 0.092 & 0.160 & 0.159 & 0.165 \\ 
  \hline
&  &ScII&TiII&VII&CrII&MnII&FeII&NiII&SrII & YII\\
\end{tabular}
\end{table}

% Abundance Table for the 3 stars redone by Marwan 25/02/2018

\newpage
\begin{deluxetable}{lccrccccr}
\tabletypesize{\normalsize}   
\tablecaption{Elemental Abundances from unblended lines for HD 30085, HD 30963, HD 174567 and HD 141556. References are defined in Tab.~5.}
\tablecolumns{7}\tablewidth{0pc}  
\tablehead{Element  & $\lambda$ (\AA) & $\log gf$ & Ref. & \multicolumn{4}{c}{$<\log(X/H)>$} \\
 \multicolumn{4}{c}{} &HD30085 & HD30963 & HD174567 & Chi Lupi A & $\sigma_{[X/H]}$ \\}
\startdata
\ion{He}{1} & 4471.470    &-2.211  &   NIST/ASD      	& -2.07 & -1.77 & -1.37  & -1.67 \\
\ion{He}{1} & 4471.474    &-0.287 &   NIST/ASD     & -2.07 & -1.77&  -1.37 & -1.67 \\
\ion{He}{1} & 4471.474    & -1.035 &  NIST/ASD      & -2.07 & -1.77& -1.37 & -1.67\\
\ion{He}{1} & 4471.485    & -1.035 & NIST/ASD	& -2.07  & -1.77& -1.37 & -1.67\\
\ion{He}{1} & 4471.489    & -0.558 & NIST/ASD	& -2.07 & -1.77 &  -1.37 & -1.67\\
\ion{He}{1} & 4471.683    & -0.910  & NIST/ASD	& -2.07 & -1.77 & -1.37 & -1.67\\
\ion{He}{1} & 5875.598    & -1.516  & NIST/ASD      &  -2.05 & -1.75 & -1.35 & -1.65 \\
\ion{He}{1} & 5875.613    & -0.339  & NIST/ASD	&  -2.05& -1.75 &  -1.35 & -1.65\\
\enddata
\label{tab:abundAll}
\end{deluxetable}

% Table for references 
%\onecolumn
\begin{deluxetable}{l|c}
\tabletypesize{\normalsize}   
\tablecaption{Bibliographical references for the atomic data.}
\tablecolumns{2}\tablewidth{12cm}  
\tablehead{Element  &   Bibliographical source \\}
\startdata
He I & Ma60=\cite{MA60} \\
He I & DR2006=\cite{DR2006} \\
C II & NIST/ASD=\cite{Kramida2017} \\
O I & KZ91=\cite{KZ91}				\\
O I & W96=\cite{W96}				\\
O I & NIST/ASD=\cite{Kramida2017}				\\
Mg II & NIST/ASD=\cite{Kramida2017}			\\
Mg II & NIST/ASD=\cite{Kramida2017}	\\
\enddata
\label{atomicbib}
\end{deluxetable}

% insert identification table for HD 30085 here

\begin{deluxetable}{lllrll}
\tabletypesize{\normalsize}  
\tablecaption{Identifications  of lines absorbing more than 2 \% for HD\,30085}
\tablecolumns{5}\tablewidth{0pc}  
\tablehead{$\lambda_{\mathrm{obs}}$ (\AA)  & $\lambda_{\mathrm{lab}}$ (\AA) & Species & $\log gf$ & $E_{low}$ & Ref. \\}
\startdata
3920.738 & 3920.627 & \ion{Fe}{2} & -1.330 & 60628.698 & NIST/VALD \\ 
3920.801 & 3920.690 & \ion{C}{2} & -0.230 & 131735.525 & NIST/VALD \\ 
3922.121 & 3922.010 & \ion{Fe}{2} & -1.110 & 73606.162 & NIST/VALD \\ 
3923.571 & 3923.460 & \ion{S}{2} & 0.440 & 130641.115 & NIST/VALD \\ 
3924.953 & 3924.842 & \ion{Fe}{2} & -1.100 & 78138.990 & NIST/VALD \\ 
3926.527 & 3926.416 & \ion{Mn}{2} & -1.600 & 55757.109 & NIST/VALD \\ 
3930.407 & 3930.296 & \ion{Fe}{1} & -1.490 & 704.007 & NIST/VALD \\ 
3930.417 & 3930.306 & \ion{Fe}{2} & -4.300 & 13671.099 & NIST/VALD \\
\enddata
\label{tab:ident}
\end{deluxetable}

%% This command is needed to show the entire author+affilation list when
%% the collaboration and author truncation commands are used.  It has to
%% go at the end of the manuscript.
\allauthors

%% Include this line if you are using the \added, \replaced, \deleted
%% commands to see a summary list of all changes at the end of the article.
\listofchanges

\end{document}